# Distributed interference management using Q-Learning in Cognitive Femtocell networks: New USRP-based Implementation

Medhat H. M. Elsayed
KINDI Center for Computing Research,
Computer Science and Engineering,
Qatar University, Qatar
Email: mhamdy@qu.edu.qa

Amr Mohamed
Computer Science and Engineering,
Qatar University, Qatar
Email: amrm@qu.edu.qa

*Abstract*—Femtocell networks have become a promising solution in supporting high data rates for 5G systems, where cell densification is performed using the small femtocells. However, femtocell networks have many challenges. One of the major challenges of femtocell networks is the interference management problem, where deployment of femtocells in the range of macro-cells may degrade the performance of the macrocell. In this paper, we develop a new platform for studying interference management in distributed femtocell networks using reinforcement learning approach. We design a complete MAC protocol to perform distributed power allocation using Q-Learning algorithm, where both independent and cooperative learning approaches are applied across network nodes. The objective of the Q-Learning algorithms is to maximize aggregate femtocells capacity, while maintaining the QoS for the Macrocell users. Furthermore, we present the realization of the algorithms using GNURadio and USRP platforms. Performance evaluation are conducted in terms of macrocell capacity convergence to a target capacity and improvement of aggregate femtocells capacity.

## I. INTRODUCTION

Cognitive Radio research has been gaining a lot of attentions in the research community for a long time. A cognitive radio is a device that is able to interact with the wireless environment and changes its radio parameters accordingly in order to achieve efficient utilization of the radio resources.

Femtocell Networks (FCN) have been proposed as a promising approach to the cell densifictaion in current LTE-A and next 5G systems, where decreasing cell size improves the spectral efficiency; hence, increases the supported data rates [1]. However, FCs face many other challenges. FCs are installed by the end-user, thus their number and positions are random and unknown a-priori. Therefore, a predefined network model cannot be identified, which complicates the centralized resource allocation across large and dense networks. Furthermore, the typical scenario for FCs is the underlay deployment, where both FCs and macrocells (MCs) share the same frequency band, hence interference management becomes a mandatory process for MC performance. As a result, distributed resource allocation algorithm is needed for improving FCs and MC capacity, while guaranteeing flexible deployment scenario.

Traditional resource management techniques can be applied for interference mitigation in FCNs. Authors in [2] proposed centralized and distributed joint power and spectrum allocation schemes, where the centralized scheme achieves better performance with more computational complexity. Authors in [3] presented downlink power control using centralized and distributed algorithms to mitigate co-channel interference. Other techniques can be found in [4], [5]. However, most of the proposed techniques either achieve high performance with high computational complexity (e.g. optimization techniques), or tend to reduce complexity and achieve reduced performance (e.g. fixed power allocation).

On the other hand, game-theoretic approaches are used for interference mitigation in FCN. A reinforcement learning approach called Q-Learning is used to automatically learn an optimal policy to maximize FCs capacity, while maintaining MC performance above certain threshold. Q-Learning offers significant advantages to achieve near-optimum decision policy through the real-time learning of the wireless environment. The power of Q-Learning approach is that it does not require any prior information about the wireless channel, where it learns the environment iteratively till it achieves the steady state after certain number of iterations. Therefore, it facilitates distributed resource allocation in FC networks. Furthermore, its computational complexity is moderate compared to other interference mitigation algorithms [6]. Q-Learning proved its significance before in both cognitive radio and WSN [7], [8], [9], [10].

On the implementation level, authors in [11] proposed RADION framework for distributed management of time-frequency resources in OFDMA-based FC networks. However, it neither considers the cross-tier interference between MC and FCs, nor the effect of FCs power control on interference mitigation. To the best of our knowledge, there is no prior implementation that considers interference mitigation using Q-Learning approach in cognitive FCNs.

The paper contributions can be summarized in the following points:

- We develop a platform using GNURadio and USRP to study the interference mitigation scenarios in cognitive femtocell networks.
- We propose two algorithms based on Q-Learning approach in order to mitigate interference and maximize aggregate femtocell capacity, while guaranteeing the macrocell capacity.
- We design a MAC protocol model to realize the distributed Q-Learning approach at the system level providing critical functions such as synchronization and SINR estimation.
- We conduct performance evaluation study to show both transient and steady state analysis of the network in real-time for power control techniques.

The paper is organized as follows: Section II presents the system model, where network model, MAC protocol and Q-Learning algorithms are discussed. Implementation details are discussed in section III. Section IV presents the performance evaluation and section V concludes the work.

## II. SYSTEM MODEL

In this section, system model for interference management in FC networks is developed. Network model, MAC protocol for the different nodes, and the Q-Learning algorithms are developed in order to facilitate the power allocation process in FC networks.

### A. Network Model

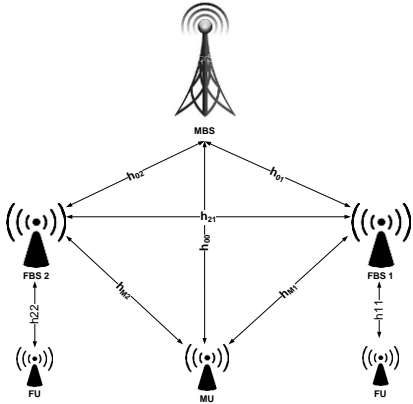

**Figure 1: Network Architecture**

We consider a wireless network consisting of one macro cell denoted by Macro Base Station (MBS) that serves one Macro User (MU). The MBS is underlaid with N FCs denoted by Femto Base Station (FBS), each serves one femto user (FU). Each node in the network (i.e. MBS, FBSs, and MU) lies in the interference range of the other nodes (Figure 1).

$p_0$, $p_m$, $p_1$, and $p_2$ denote the transmission powers of the MBS, MU, FBS1, and FBS2. In this model, we consider only power allocation for the FBSs and a specified number of subchannels for each node, hence each FBS has (k) subchannels for its Femto Users (FUs). The performance of the interference management model is analyzed in terms of the capacity of the nodes measured in (bits/sec/Hz):

$$C_n = \log_2(1+\Gamma_n), C_m = \log_2(1+\Gamma_m), C_0 = \sum_{i=1}^{n} C_i \quad (1)$$

Where, $C_n$ denotes the aggregate capacity of FBS n over all k-subchannels; $C_m$ is the aggregate MBS capacity over all k-subchannels; $C_0$ is the aggregate network capacity for all FBSs over all k-subchannels; $\gamma_n$ and $\gamma_m$: are the Signal to Interference plus Noise Ratio (SINR) of FBSs and MBS, respectively. For each node, the SINR can be expressed as follows:

$$\Gamma_1 = \frac{P_1 h_{01}}{N_{AWGN} + P_m h_{m1} + P_0 h_{01} + P_2 h_{21} + ... + P_n h_{n1}} \quad (2a)$$

$$\Gamma_m = \frac{P_0 h_{00}}{N_{AWGN} + P_1 h_{01} + P_2 h_{02} + ... + P_n h_{0n}} \quad (2b)$$

where, $P_0$, $P_N$ and $P_m$ are the transmission powers for the MBS, FBSN, MU, respectively; $h_{00}$, $h_{01}$, $h_{02}$, $h_{21}$, $h_{m1}$, and $h_{m2}$ are the channel gains between MBS and MU, MBS and FBS1, MBS and FBS1, FBS1 and FBS2, MU and FBS1, MU and FBS2, respectively; and $N_{AWGN}$ is the Additive white Gaussian noise (AWGN). Channel gains estimation, hence the SINRs, are performed periodically to maintain updated channel state information (CSI). This estimation is performed for each subchannel, where 2-subchannels are used in our model for proof-of-concept implementation.

### B. Power Allocation using Q-Learning: MAC protocol

The CSI (i.e. $\gamma_n$ and $\gamma_m$) are utilized by the Q-Learning algorithms to perform the power allocation process for each FBS. Therefore, a communication MAC protocol is needed to facilitate the process of acquiring these CSI parameters.

Power Allocation using Q-Learning (PAQ) MAC protocol facilitates nodes' synchronization, SINR estimation, incremental node deployment, and distributed and cooperative power allocation using Q-Learning algorithm. We build on our previous work in [11], where the testbed implementation is utilized for the development of PAQ MAC protocol. PAQ frame structure is divided into three states: Sync/Re-Sync, Acquisition, and Q-Learning Power Allocation (QPA) states (Figure 2).

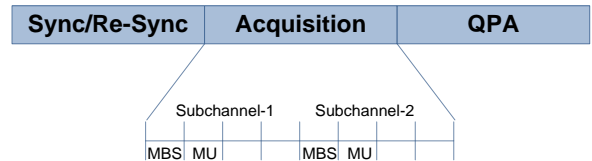

**Figure 2: Frame structure for the DPA-Q MAC protocol**

**Sync/Re-Sync State:** In this state, MBS performs the synchronization through broadcasting a beacon packet periodically to compensate for any misalignment in clock frequencies of each node. Furthermore, the beacon packet includes some information about the network such as the number of time slots in each state, and the time slot occupancy in the following states.

**Acquisition State:** The acquisition state is designed to be TDMA-based state, where each node broadcasts its transmission power to all other nodes. The acquisition state is further sub-divided into two 4-time slot sub-states, where each sub-state is devoted for specific subchannel. in each sub-state, the CSI is estimated to find the channel gains of all links shown in Figure 1 for each subchannel. For instance, for sub-state 1, FBS1 may broadcast its transmission power in the third time slot, while at the same time the other nodes will estimate the channel gains with FBS1 (i.e. $h_{01}$, $h_{m1}$, and $h_{21}$) on subchannel 1. The same applies to the second sub-state, which is used for CSI estimation on subchannel 2.

The first and second time slots in the acquisition state is reserved to MBS, and MU, respectively. To ensure incremental deployment in the network, slotted aloha multiple access protocol is used for new FBS entering the network. We anticipate good performance of the slotted aloha protocol as this scenario considers low number of FBSs [12].

**QPA state:** The Q-Learning Power Allocation (QPA) state is devoted for each FBS to perform the Q-Learning algorithm as will be discussed in II-C. In each QPA time slot, each node calculates its capacity (i.e. using equation (1)) and broadcasts it to the other nodes. The Q-Learning algorithm is performed at the end of the QPA state to allocate new power actions to each FBS. Such actions are used as transmission power actions by the FBSs in the subsequent acquisition state.

*C. Q-Learning Algorithms*

In this subsection, Q-Learning power allocation algorithms are presented. Q-Learning facilitates the distributed power allocation, where each FBS is modeled as an agent that learns the CSI of the network and performs certain actions accordingly [13]. Two Q-learning paradigms exist; Independent Learning (IL) and Cooperative Learning (CL). In the IL paradigm, nodes allocate actions independently, while in CL paradigm, nodes share their allocation policies to achieve better performance. It is intuitive that the CL approach achieves better performance than the IL approach. However, CL approach adds communication overhead that is stemmed from sharing Q-Learning parameters. Therefore, CL approach gives better performance on the price of more communication overhead. Our implementation model considers both the IL and CL approaches for performance comparison on a real-time testbed.

*1) Partial Distributed Power Allocation using Q-Learning (PDPA-Q)*

The PDPA-Q algorithm considers the IL approach, where each node performs the Q-Learning algorithm to allocate transmission power based on the acquired CSI. The objective is to maximize aggregate FBSs capacity (i.e. $C_0$ defined in equation (1)) through transmission power allocation, while not affecting MBS target capacity (i.e. managing the interference on the MBS).

The network is modeled as a set of agents playing a game; Each agent takes an action to play, and a corresponding reward is given to the agent. In our network model, agents are the FBSs; actions are the quantized transmission power levels FBSs can allocate. The power action of each FBS will affect the CSI of the network, hence a reward/cost should be given to each FBS accordingly. Q-Learning network model follows:

**Agents:** FBSs.

**States:** The state is defined as \{I\}, where \{I\} indicates the level of interference measured at the MBS.

$$I = \begin{cases} 1 & C_m < B_0 \\ 0 & C_m \geq B_0 \end{cases} \quad (3)$$

Where, $B_0$ is the target capacity; and $C_m$ is the MBS capacity.

**Actions:** The set of actions is a vector of power level pairs that each FBS can allocate on its subchannels. In our model, there are 2 subchannels for each FBS, hence the allocated action will be a vector of two power levels for the two subchannels.

**Reward Function:** FBS reward in accordance to state-action pair over all subchannels is:

$$R = e^{-(C_m - B_0)^2} - e^{-C_n} \quad (4)$$

Where, $C_n$ is the capacity of FBS n [13]. PDPA-Q assigns each FBS a Q-Table, where each state-action pair takes a Q-value $Q(s, a)$. The Q-value is defined to be the expected discounted reward over an infinite time when action $a_l$ is performed in state $s_m$. Table 2 presents the algorithm steps.

| PDPA-Q learning algorithm |
|---|
| 0. Initialize the Q-Table of each FBS with zero values for all Q-Values |
| 1. Detect the environment current state $s_{m+1}$ |
| 2. Update Reward (R) at the current FBS state $s_m$ Note that FBS is still at its current state $s_m$ |
| 3. Update Q-value at the current FBS state $s_m$ and current action $a_l$ as follows: $Q(s_m, a_l) = (1 - \alpha)Q(s_{m-1}, a_l) + \alpha[R + \gamma \, max(Q(s_{m-1}, a))]$ (5) |
| 4. Switch to the new state $s_{m+1}$ |
| 5. Choose action that maximize Q-value at the new state $s_{m+1}$ |
| 6. Go back to step 1 |

**Table 1: DPA-Q LEARNING ALGORITHM**

After acquiring the CSI in the acquisition state, the PDPA-Q learning algorithm is performed in the QPA state for each FBS. The CSI is affected by the new power allocation performed in the QPA. Therefore, the acquisition state is important in each frame in order to get updated CSI to be used in the PDPA-Q algorithm. It is important to note that the PDPA-Q algorithm does not consider the cooperation between FBSs, hence the power allocation process is performed independently by each FBS.

*2) Cooperative Distributed Power Allocation using Q-Learning (CDPA-Q)*

The CDPA-Q approach follows the CL paradigm, where the Q-row (i.e. the row of the Q-Table corresponding to the current state) of each FBS is shared among the FBSs in order to improve the decision process. The CDPA-Q algorithm follows the same algorithm as PDPA-Q except for the action allocation method. In this case, the Q-row corresponding to the current state of each FBS is shared among the FBS, and the action selection is performed according to the action that maximizes the summation of the Q-values of all the FBS. This is referred to as maximizing the global Q-value.

$$Global\ Q - Value = \sum_{1}^{n}(Q(s,a)) \qquad (6)$$

Where (n) is the number of FBS in the network. The equation of the global Q-value (equation 6) is a vector summation of the Q-values in the respective current state of each FBS. The action selection depends on the maximum global Q-value. It can be observed that the CL approach results in a global action selection, where all the FBSs will select the same action which may degrades its performance.

## III. Q-LEARNING IMPLEMENTATION USING USRP PLATFORM

Most of the communication systems algorithms lack the realization on a hardware platform. The Q-Learning algorithms presented provide low computational complexity, and distributed implementation on each node. In this work, we utilize the USRP platform for implementing the PDPA-Q and CDPA-Q algorithms. USRP is a very efficient platform for designing flexible solutions, and very helpful in proof-of-concept designs. Each node is composed of two parts: a RF open-source reconfigurable front-end to support the RF communication, and a software platform to perform the DSP processing. The RF front-end is implemented with the USRP N210 [14] with SBX Rx/Tx [15], which operates in the band of 400-4400 MHz and provides up to 100 mW output power. Software modules (i.e. physical and MAC layers) are implemented using GNURadio SDR platform. Figure 3 presents a conceptual model for MBS, MU, FBS nodes. Both MBS and MU nodes have the same hardware model.

*1) Q-Learning MAC Protocol Block*

The MAC block is a common block in all network nodes, where it performs the MAC functionality discussed in section II-B.

*2) Q-Learning MAC algorithm Block*

This block refers to the implementation of the Q-Learning algorithms discussed in section II-C. This block is only implemented in the FBS node, where it is the only node that performs the Q-learning algorithms in the network.

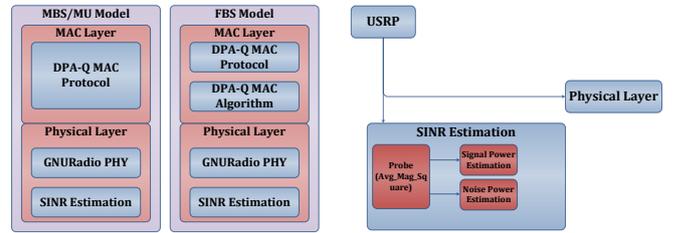

**Figure 3: Conceptual Model for MBS, MU and FBS nodes**

**Figure 4: SINR Estimation Block Diagram**

*3) GNURadio PHY Block*

This block performs the physical layer functionality, where it is mainly supported by the physical layer modules in the GNURadio software.

*4) SINR Estimation Block*

USRP power estimation constitute an important problem because of the inconsistent results coming from the estimation blocks in the USRP. In the proposed Q-Learning algorithms, SINR estimation is mandatory process for the calculation of node network capacity. The SINR estimation process is performed using the probe block in GNURadio (Figure 4), where the probe performs an average magnitude square process on the samples acquired during the sensing time. The acquired results from the probe block is further used to estimate the signal and noise powers.

Signal power estimation is performed by averaging the acquired samples from the probe block over certain time. For instance, we collect 10 outcomes from the probe block (i.e., note that each outcome represents 10 msecond sensing), then signal power is the mean of those 10 outcomes. On the other hand, noise power is estimated by considering the variance of the collected outcomes. This improves the power and noise estimation.

## IV. Q-LEARNING PERFORMANCE EVALUATION

In this section, performance evaluation of the proposed Q-learning algorithms is conducted using the USRP N210 platform. Performance measures are the convergence of MBS capacity to the target capacity, and the convergence and improvement of FBSs aggregate capacity compared to other power allocation schemes.

*A. Implementation Setup*

We consider the network model presented in Figure 1, where one MBS, one MU, and two FBSs are represented by four USRPs. Table 2 lists all the parameters used for the USRP implementation setup.

| Subchannel frequencies | |
|---|---|
| Common Control Channel | 2.401624512GHz |
| Subchannel frequency 1 | 2.402124512GHz |
| Subchannel frequency 2 | 2.402624512GHz |
| Q-Learning MAC protocol parameters | |
| Acquisition state time slots | 2 sub-states with 4 time slots each. |
| QPA state time slots | 4 |
| Time slot duration | 0.5 second |
| Sensing duration | 10 mseconds |
| Modulation scheme | gmsk |
| Data Rate | 0.5Mbps |
| Q-learning algorithm parameters | |
| $B_0$ (target capacity) | 11 bps/Hz |
| $\alpha$ (learning rate) | 0.5 |
| $\gamma$ | 0.9 |
| FBSs Actions (power levels) | 0:5:30 dB |
| MBS transmission power level | 20 dB |

**Table 2: Q-Learning USRP implementation parameters**

*B. Performance Evaluation Results*

In this subsection, we show the performance evaluation of the two Q-Learning algorithms applied on our USRP platform. Furthermore, Equal Power (EP) allocation has also been simulated for comparison with the Q-Learning algorithms presented. In our simulation, each FBS has two subchannels to use as mentioned in Table 1. In this case, a power level allocation is required for each subchannel, where finding a combined near-optimum subchannel power allocation for the two FBSs would require exhaustive search of $((7^2)^2 = 2401)$ iterations, where 7 power levels are used.

Simulation scenarios considered are: simulation using one FBS in the interference range of the MBS, simulation using two FBSs in the interference range of the MBS, and incremental deployment of two FBSs.

Figure 5 and Figure 6 show the MBS and FBSs aggregate capacity for one FBS, respectively. It is obvious that the EP algorithm achieves high performance for the MBS, while the FBS performance is low. Both of the PDPA-Q and CDPA-Q increases the performance of the FBS, while maintaining the MBS performance above the target capacity.

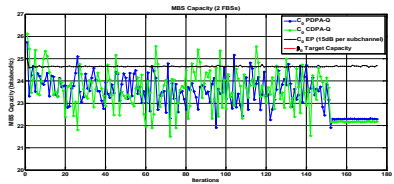

Figure **7** and Figure 8 show the capacity of the MBS, and the aggregate capacity for two FBSs, respectively. Four observations can be drawn from the figures; First, the PDPA-Q and CDPA-Q algorithms are compared to the EP allocation algorithm, where the EP algorithm achieves better performance for the MBS but lower performance for the aggregate FBSs capacity. Second, both the PDPA-Q and CDPA-Q algorithms maximize the performance of the FBSs while guaranteeing near target capacity for the MBS. Third, the CDPA-Q further maximizes the performance of the FBSs, and further approaches the target capacity for the MBS. And finally, figures show that the Q-Learning algorithms can achieve the near-optimum performance with much lower iterations than the exhaustive search algorithm.

Figure 9 and Figure 10 show the performance for the incremental deployment case. Incremental deployment can take two forms according the arrival of the second FBS. Either FBS 2 enters the network within the exploration phase of FBS 1, or after FBS 1 completes the exploration phase, which is the worst case. The figures simulate the worst case, where it can be shown that the FBS performance increases with same MBS performance, which is above the target capacity.

V. CONCLUSION

In this paper, interference management in femtocell networks using Q-Learning approach has been addressed using GNURadio and USRP platforms. Power allocation using Q-Learning approach has been conducted. The contributions of the paper can be summarized in the following point. First, Q-Learning algorithm has been adopted to mitigate interference and maximize aggregate FBSs capacity, while guaranteeing MBS performance. Second, a PAQ MAC protocol has been designed to facilitate the distributed Q-Learning approach through supporting critical functions such as synchronization and SINR estimation. Third, a testbed has been implemented for acting the interference scenarios in femtocell networks. The performance study showed both transient and steady state analysis of the network in real-time. Furthermore, a comparison between the proposed approach and equal power allocation approach revealed the advantages of the Q-Learning approach in real-time scenarios.

VI. ACKNOWLEDGMENT

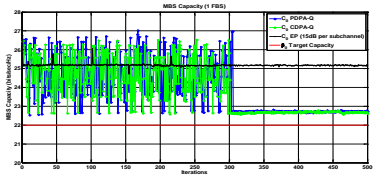

**Figure 5: MBS Capacity (one MU, and one FBS)**

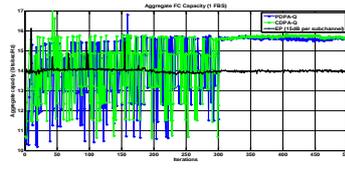

**Figure 6: Aggregate FCs Capactiy (one MU, and one FBSs)**

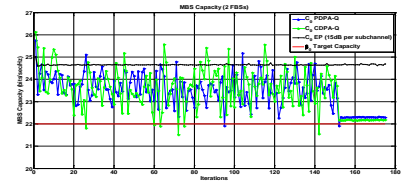

**Figure 7: MBS Capacity (one MU, and two FBSs)**

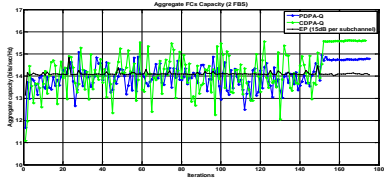

**Figure 8: Aggregate FCs Capactiy (one MU, and two FBSs)**

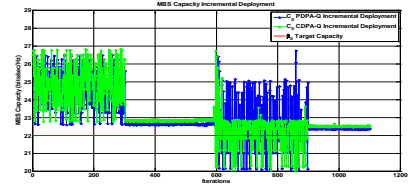

**Figure 9: MBS Capacity (Incremental Deployment)**

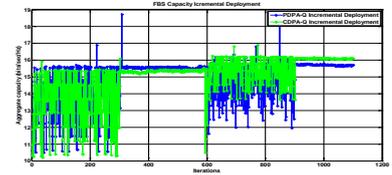

**Figure 10: Aggregate FCs Capactiy (Incremental Deployment)**